\documentclass[12pt]{article}
\usepackage{amssymb}
\usepackage{amsmath}

\begin{document}

\begin{titlepage}
\title{\bf Hamiltonian Mechanics on Quaternion K\"{a}hler Manifolds}
\author{ Mehmet Tekkoyun \footnote{Corresponding author. E-mail address: tekkoyun@pau.edu.tr; Tel: +902582953616; Fax: +902582953593}\\
{\small Department of Mathematics, Pamukkale University,}\\
{\small 20070 Denizli, Turkey}}
\date{\today}
\maketitle

\begin{abstract}

The goal of this study is  to present quaternion K\"{a}hler
analogue of Hamiltonian mechanics. Finally, the some results related
to quaternion K\"{a}hler dynamical systems were also given.

{\bf Keywords:} Quaternion K\"{a}hler Geometry, Hamiltonian
Mechanics.

{\bf MSC:} 53C15; 70H05.

\end{abstract}
\end{titlepage}

\section{Introduction}

It is well-known that modern differential geometry explains explicitly the
dynamics of Hamiltonians. Therefore, if $Q$ is an $m$-dimensional
configuration manifold and $\mathbf{H}:T^{\ast }Q\rightarrow \mathbf{R}$%
\textbf{\ }is a regular Hamiltonian function, then there is a unique vector
field $X$ on $T^{\ast }Q$ such that dynamic equations are given by
\begin{equation}
\,\,i_{X}\Phi =d\mathbf{H}  \label{1.1}
\end{equation}%
where $\Phi $ indicates the symplectic form. The triple $(T^{\ast }Q,\Phi
,X) $ are called \textit{Hamiltonian system }on the cotangent bundle $%
T^{\ast }Q. $

Nowadays, there are a lot of studies about Hamiltonian mechanics,
formalisms, systems and equations \cite{deleon, tekkoyun} and there in.
There are real, complex, paracomplex and other analogues. We say that in
order to obtain different analogous in different spaces is possible.

Quaternions were invented by Sir William Rowan Hamiltonian as an extension
to the complex numbers. Hamiltonian's defining relation is most succinctly
written as:

\begin{equation}
i^{2}=j^{2}=k^{2}=ijk=-1  \label{1.2}
\end{equation}%
If it is compared to the calculus of vectors, quaternions have slipped into
the realm of obscurity. They do however still find use in the computation of
rotations. A lot of physical laws in classical, relativistic, and quantum
mechanics can be written pleasantly by means of quaternions. Some physicists
hope they will find deeper understanding of the universe by restating basic
principles in terms of quaternion algebra. It is well-known that quaternions
are useful for representing rotations in both quantum and classical
mechanics \cite{dan} .

In this study, therefore, Hamiltonian equations related to mechanical
systems on quaternion K\"{a}hler manifold have been presented.

\section{Preliminaries}

Throughout this paper, all mathematical objects and mappings are assumed to
be smooth, i.e. infinitely differentiable and Einstein convention of
summarizing is adopted. $\mathcal{F}(M)$, $\chi (M)$ and $\Lambda ^{1}(M)$
denote the set of functions on $M$, the set of vector fields on $M$ and the
set of 1-forms on $M$, respectively.

\subsection{Quaternion K\"{a}hler Manifolds}

Let $M$ be an n-dimensional manifold. It has a 3-dimensional vector bundle $%
V $ consisting of tensors of type (1,1). The manifold $M$ satisfies the
condition given by:

(a) In any coordinate neighborhood $U$ of $M$, there exists a local basis $%
\{F,G,H\}$ of $V$ such that

\begin{equation}
F^{2}=G^{2}=\text{ }H^{2}=FGH=-I.  \label{2.1}
\end{equation}%
$I$ denotes the identity tensor of type (1,1) in $M$. Such a local basis $%
\{F,G,H\}$ is called a canonical local basis of the bundle $V$ in $U$. Then $%
V$ is said to be an almost quaternion structure in $M$, and $M$ with $V$ is
an almost quaternion manifold denoted by $(M,V)$. An almost quaternion
manifold $M$ is of dimension $n=4m$ $(m\geqslant 1).$ In any almost
quaternion manifold $(M,V)$, there is a Riemannian metric tensor field $g$
such that
\begin{equation}
g(\phi X,Y)+g(X,\phi Y)=0  \label{2.2}
\end{equation}%
for any cross-section $\phi $ and any vector fields $X,Y$ of $M.$An almost
quaternion structure $V$ fixed with a Riemannian metric $g$ is called an
almost quaternion metric structure$.$ A manifold $M$ endowed with an almost
quaternion metric structure $\{g,V\}$ is said to be an almost quaternion
metric manifold denoted by $(M,g,V).$ Let $\{F,G,H\}$ be a canonical local
basis of $V$ an almost quaternion manifold $(M,g,V)$. Since each of $F,G$
and $H$ is almost Hermitian with respect to $g$, setting

\begin{equation}
\Phi (X,Y)=g(FX,Y),~\text{\ }\Psi (X,Y)=g(GX,Y),\text{~}\Theta (X,Y)=g(HX,Y)
\label{2.3}
\end{equation}

for any vector fields $X$ and $Y$, we see that $\Phi ,\Psi $ and $\Theta $
are local 2-forms.

Assume that the Riemannian connection $\nabla $ of $(M,g,V)$ satisfies the
conditions as follows:

$(b)$ If $\phi $ is a cross-section (local or global) of the bundle $V$,
then $V_{X}\phi $ is also a cross-section of $V$,\ where $X$ is an arbitrary
vector field in $M$. From (\ref{2.1}) we see that the condition $(b)$ is
equivalent to the following condition:

$(b^{\prime })$ If $F,G,H$ is a canonical local basis of $V$, then
\begin{equation}
\nabla _{X}F=r(X)G-q(X)H,\text{ \ }\nabla _{X}G=-r(X)F+p(X)H,\text{ \ }%
\nabla _{X}H=q(X)F-p(X)G  \label{2.4}
\end{equation}

for any vector field $X$, where $p,q$ and $r$ are certain local 1-forms. If
an almost quaternion metric manifold $M$ satisfies the condition $(b)$ or $%
(b^{\prime })$, then $M$ is said to be a quaternion K\"{a}hler manifold and
an almost quaternion structure of $M$ is called a quaternion K\"{a}hler
structure. \cite{yano}

Let $\left\{ x_{i},x_{n+i},x_{2n+i},x_{3n+i}\right\} ,$ $i=\overline{1,n}$
be a real coordinate system on a neighborhood $U$ of $M,$ and suppose that
let $\left\{ \frac{\partial }{\partial x_{i}},\frac{\partial }{\partial
x_{n+i}},\frac{\partial }{\partial x_{2n+i}},\frac{\partial }{\partial
x_{3n+i}}\right\} $ and $\{dx_{i},dx_{n+i},dx_{2n+i},dx_{3n+i}\}$ be natural
bases over $\mathbf{R}$ of the tangent space $T(M)$ and the cotangent space $%
T^{\ast }(M)$ of $M,$ respectively$.$Taking into \cite{burdujan}, then the
following expression can be found%
\begin{eqnarray}
F(\frac{\partial }{\partial x_{i}}) &=&\frac{\partial }{\partial x_{n+i}},%
\text{ }F(\frac{\partial }{\partial x_{n+i}})=-\frac{\partial }{\partial
x_{i}},\text{ }F(\frac{\partial }{\partial x_{2n+i}})=\frac{\partial }{%
\partial x_{3n+i}},\text{ }F(\frac{\partial }{\partial x_{3n+i}})=-\frac{%
\partial }{\partial x_{2n+i}}  \label{2.5} \\
G(\frac{\partial }{\partial x_{i}}) &=&\frac{\partial }{\partial x_{2n+i}},%
\text{ }G(\frac{\partial }{\partial x_{n+i}})=-\frac{\partial }{\partial
x_{3n+i}},\text{ }G(\frac{\partial }{\partial x_{2n+i}})=-\frac{\partial }{%
\partial x_{i}},\text{ }G(\frac{\partial }{\partial x_{3n+i}})=\frac{%
\partial }{\partial x_{n+i}}  \notag \\
H(\frac{\partial }{\partial x_{i}}) &=&\frac{\partial }{\partial x_{3n+i}},%
\text{ }H(\frac{\partial }{\partial x_{n+i}})=\frac{\partial }{\partial
x_{2n+i}},\text{ }H(\frac{\partial }{\partial x_{2n+i}})=-\frac{\partial }{%
\partial x_{n+i}},\text{ }H(\frac{\partial }{\partial x_{3n+i}})=-\frac{%
\partial }{\partial x_{i}}  \notag
\end{eqnarray}

A canonical local basis$\{F^{\ast },G^{\ast },H^{\ast }\}$ of $V^{\ast }$ of
the cotangent space $T^{\ast }(M)$ of manifold $M$ satisfies the condition
as follows:

\begin{equation}
F^{\ast 2}=G^{^{\ast }2}=\text{ }H^{\ast 2}=F^{\ast }G^{\ast }H^{\ast }=-I,
\label{2.6}
\end{equation}
defining by

\begin{eqnarray*}
F^{\ast }(dx_{i}) &=&dx_{n+i},\text{ }F^{\ast }(dx_{n+i})=-dx_{i},\text{ }%
F^{\ast }(dx_{2n+i})=dx_{3n+i},\text{ }F^{\ast }(dx_{3n+i})=-dx_{2n+i}, \\
G^{\ast }(dx_{i}) &=&dx_{2n+i},\text{ }G^{\ast }(dx_{n+i})=-dx_{3n+i},\text{
}G^{\ast }(dx_{2n+i})=-dx_{i},\text{ }G^{\ast }(dx_{3n+i})=dx_{n+i}, \\
H^{\ast }(dx_{i}) &=&dx_{3n+i},\text{ }H^{\ast }(dx_{n+i})=dx_{2n+i},\text{ }%
H^{\ast }(dx_{2n+i})=-dx_{n+i},\text{ }H^{\ast }(dx_{3n+i})=-dx_{i}.
\end{eqnarray*}

\section{Hamiltonian Mechanics}

Here, we obtain Hamiltonian equations and Hamiltonian mechanical system for
quantum and classical mechanics structured on quaternion K\"{a}hler manifold
$(M,V).$

Firstly, let $(M,V)$ be a quaternion K\"{a}hler manifold. Assume that a
component of almost quaternion structure $V^{\ast }$, a Liouville form and a
1-form on quaternion K\"{a}hler manifold $(M,V)$ are shown by $F^{\ast }$, $%
\lambda _{F^{\ast }}$ and $\omega _{F^{\ast }}$, respectively$.$

Then $\omega _{F^{\ast }}=\frac{1}{2}%
(x_{i}dx_{i}+x_{n+i}dx_{n+i}+x_{2n+i}dx_{2n+i}+x_{3n+i}dx_{3n+i})$ and $%
\lambda _{F^{\ast }}=F^{\ast }(\omega _{F^{\ast }})=\frac{1}{2}%
(x_{i}dx_{n+i}-x_{n+i}dx_{i}+x_{2n+i}dx_{3n+i}-x_{3n+i}dx_{2n+i}).$ It is
concluded that if $\Phi _{F^{\ast }}$ is a closed K\"{a}hler form on
quaternion K\"{a}hler manifold $(M,V),$ then $\Phi _{F^{\ast }}$ is also a
symplectic structure on quaternion K\"{a}hler manifold $(M,V)$.

Consider that Hamilton vector field $X$ associated with Hamiltonian energy $%
\mathbf{H}$ is given by
\begin{equation}
X=X^{i}\frac{\partial }{\partial x_{i}}+X^{n+i}\frac{\partial }{\partial
x_{n+i}}+X^{2n+i}\frac{\partial }{\partial x_{2n+i}}+X^{3n+i}\frac{\partial
}{\partial x_{3n+i}}.  \label{4.2}
\end{equation}

Then
\begin{equation*}
\Phi _{F^{\ast }}=-d\lambda _{F^{\ast }}=dx_{n+i}\wedge
dx_{i}+dx_{3n+i}\wedge dx_{2n+i}
\end{equation*}%
and
\begin{equation}
i_{X}\Phi _{F^{\ast }}=\Phi _{F^{\ast
}}(X)=X^{n+i}dx_{i}-X^{i}dx_{n+i}+X^{3n+i}dx_{2n+i}-X^{2n+i}dx_{3n+i}.
\label{4.4}
\end{equation}%
Moreover, the differential of Hamiltonian energy is obtained as follows:
\begin{equation}
d\mathbf{H}=\frac{\partial \mathbf{H}}{\partial x_{i}}dx_{i}+\frac{\partial
\mathbf{H}}{\partial x_{n+i}}dx_{n+i}+\frac{\partial \mathbf{H}}{\partial
x_{2n+i}}dx_{2n+i}+\frac{\partial \mathbf{H}}{\partial x_{3n+i}}dx_{3n+i}.
\label{4.5}
\end{equation}%
According to \textbf{Eq.}(\ref{1.1}), if equaled \textbf{Eq. }(\ref{4.4})
and \textbf{Eq. }(\ref{4.5}), the Hamiltonian vector field is found as
follows:
\begin{equation}
X=-\frac{\partial \mathbf{H}}{\partial x_{n+i}}\frac{\partial }{\partial
x_{i}}+\frac{\partial \mathbf{H}}{\partial x_{i}}\frac{\partial }{\partial
x_{n+i}}-\frac{\partial \mathbf{H}}{\partial x_{3n+i}}\frac{\partial }{%
\partial x_{2n+i}}+\frac{\partial \mathbf{H}}{\partial x_{2n+i}}\frac{%
\partial }{\partial x_{3n+i}}.  \label{4.6}
\end{equation}

Suppose that a curve
\begin{equation}
\alpha :I\subset \mathbf{R}\rightarrow M  \label{4.7}
\end{equation}%
be an integral curve of the Hamiltonian vector field $X$, i.e.,
\begin{equation}
X(\alpha (t))=\overset{.}{\alpha },\,\,t\in I.  \label{4.8}
\end{equation}%
In the local coordinates, it is obtained that
\begin{equation}
\alpha (t)=(x_{i},x_{n+i},x_{2n+i},x_{3n+i})  \label{4.9}
\end{equation}%
and
\begin{equation}
\overset{.}{\alpha }(t)=\frac{dx_{i}}{dt}\frac{\partial }{\partial x_{i}}+%
\frac{dx_{n+i}}{dt}\frac{\partial }{\partial x_{n+i}}+\frac{dx_{2n+i}}{dt}%
\frac{\partial }{\partial x_{2n+i}}+\frac{dx_{3n+i}}{dt}\frac{\partial }{%
\partial x_{3n+i}}.  \label{4.10}
\end{equation}
Considering \textbf{Eq. }(\ref{4.8}), if equaled \textbf{Eq. }(\ref{4.6}) and%
\textbf{\ Eq. }(\ref{4.10}), it follows
\begin{equation}
\frac{dx_{i}}{dt}=-\frac{\partial \mathbf{H}}{\partial x_{n+i}},\text{ }%
\frac{dx_{n+i}}{dt}=\frac{\partial \mathbf{H}}{\partial x_{i}},\text{ }\frac{%
dx_{2n+i}}{dt}=-\frac{\partial \mathbf{H}}{\partial x_{3n+i}},\text{ }\frac{%
dx_{3n+i}}{dt}=\frac{\partial \mathbf{H}}{\partial x_{2n+i}}  \label{4.11}
\end{equation}%
Thus, the equations obtained in \textbf{Eq. }(\ref{4.11}) are seen to be
\textit{Hamiltonian equations} with respect to component $F^{\ast }$ of
almost quaternion structure $V^{\ast }$ on quaternion K\"{a}hler manifold $%
(M,V),$ and then the triple $(M,\Phi _{F^{\ast }},X)$ is seen to be a
\textit{Hamiltonian mechanical system }on quaternion K\"{a}hler manifold $%
(M,V)$.

Secondly, let $(M,V)$ be a quaternion K\"{a}hler manifold. Suppose that an
element of almost quaternion structure $V^{\ast }$, a Liouville form and a
1-form on quaternion K\"{a}hler manifold $(M,V)$ are denoted by $G^{\ast }$,
$\lambda _{G^{\ast }}$ and $\omega _{G^{\ast }}$, respectively$.$

Then $\omega _{G^{\ast }}=\frac{1}{2}%
(x_{i}dx_{i}+x_{n+i}dx_{n+i}+x_{2n+i}dx_{2n+i}+x_{3n+i}dx_{3n+i})$ and $%
\lambda _{G^{\ast }}=G^{\ast }(\omega _{G^{\ast }})=\frac{1}{2}%
(x_{i}dx_{2n+i}-x_{n+i}dx_{3n+i}-x_{2n+i}dx_{i}+x_{3n+i}dx_{n+i}).$ It is
known if $\Phi _{G^{\ast }}$ is a closed K\"{a}hler form on quaternion K\"{a}%
hler manifold $(M,V),$ then $\Phi _{G^{\ast }}$ is also a symplectic
structure on quaternion K\"{a}hler manifold $(M,V)$.

Let $X$ be a Hamiltonian vector field related to Hamiltonian energy $\mathbf{%
H}$ and given by \textbf{Eq. }(\ref{4.2}).

Considering
\begin{equation}
\Phi _{G^{\ast }}=-d\lambda _{G^{\ast }}=dx_{2n+i}\wedge
dx_{i}+dx_{n+i}\wedge dx_{3n+i},  \label{4.12}
\end{equation}%
then we calculate%
\begin{equation}
i_{X}\Phi _{G^{\ast }}=\Phi _{G^{\ast }}(X)=X^{2n+i}dx_{i}-X^{i}\frac{%
\partial }{\partial x_{i}}dx_{2n+i}+X^{n+i}dx_{3n+i}-X^{3n+i}dx_{n+i}.
\label{4.13}
\end{equation}%
Besides, the differential of Hamiltonian energy is as follows:
\begin{equation}
d\mathbf{H}=\frac{\partial \mathbf{H}}{\partial x_{i}}dx_{i}+\frac{\partial
\mathbf{H}}{\partial x_{n+i}}dx_{n+i}+\frac{\partial \mathbf{H}}{\partial
x_{2n+i}}dx_{2n+i}+\frac{\partial \mathbf{H}}{\partial x_{3n+i}}dx_{3n+i}.
\label{4.14}
\end{equation}%
According to \textbf{Eq.}(\ref{1.1}), if we equal \textbf{Eq. }(\ref{4.13})
and \textbf{Eq. }(\ref{4.14}), it follows
\begin{equation}
X=-\frac{\partial \mathbf{H}}{\partial x_{2n+i}}\frac{\partial }{\partial
x_{i}}+\frac{\partial \mathbf{H}}{\partial x_{3n+i}}\frac{\partial }{%
\partial x_{n+i}}+\frac{\partial \mathbf{H}}{\partial x_{i}}\frac{\partial }{%
\partial x_{2n+i}}-\frac{\partial \mathbf{H}}{\partial x_{n+i}}\frac{%
\partial }{\partial x_{3n+i}}.  \label{4.15}
\end{equation}

Considering \textbf{Eq. }(\ref{4.8}), \textbf{Eqs. }(\ref{4.10}) and\textbf{%
\ }(\ref{4.15}) are equal, we find equations
\begin{equation}
\frac{dx_{i}}{dt}=-\frac{\partial \mathbf{H}}{\partial x_{2n+i}},\text{ }%
\frac{dx_{n+i}}{dt}=\frac{\partial \mathbf{H}}{\partial x_{3n+i}},\text{ }%
\frac{dx_{2n+i}}{dt}=\frac{\partial \mathbf{H}}{\partial x_{i}},\text{ }%
\frac{dx_{3n+i}}{dt}=-\frac{\partial \mathbf{H}}{\partial x_{n+i}}
\label{4.16}
\end{equation}%
In the end, the equations obtained in \textbf{Eq. }(\ref{4.16}) are known to
be \textit{Hamiltonian equations} with respect to component $G^{\ast }$ of
almost quaternion structure $V^{\ast }$ on quaternion K\"{a}hler manifold $%
(M,V),$ and then the triple $(M,\Phi _{G^{\ast }},X)$ is a \textit{%
Hamiltonian mechanical system }on quaternion K\"{a}hler manifold $(M,V)$.

Thirdly, let $(M,V)$ be a quaternion K\"{a}hler manifold. By $H^{\ast }$, $%
\lambda _{H^{\ast }}$ and $\omega _{H^{\ast }},$ we denote a component of
almost quaternion structure $V^{\ast }$, a Liouville form and a 1-form on
quaternion K\"{a}hler manifold $(M,V)$, respectively$.$

Then $\omega _{H^{\ast }}=\frac{1}{2}%
(x_{i}dx_{i}+x_{n+i}dx_{n+i}+x_{2n+i}dx_{2n+i}+x_{3n+i}dx_{3n+i})$ and $%
\lambda _{H^{\ast }}=H^{\ast }(\omega _{H^{\ast }})=\frac{1}{2}%
(x_{i}dx_{3n+i}+x_{n+i}dx_{2n+i}-x_{2n+i}dx_{n+i}-x_{3n+i}dx_{i}).$ It is
well-known that if $\Phi _{H^{\ast }}$ is a closed K\"{a}hler form on
quaternion K\"{a}hler manifold $(M,V),$ then $\Phi _{H^{\ast }}$ is also a
symplectic structure on quaternion K\"{a}hler manifold $(M,V)$.

Consider $X$ . It is Hamiltonian vector field connected with Hamiltonian
energy $\mathbf{H}$ and given by \textbf{Eq. }(\ref{4.2}).

Taking into
\begin{equation}
\Phi _{H^{\ast }}=-d\lambda _{H^{\ast }}=dx_{3n+i}\wedge
dx_{i}+dx_{2n+i}\wedge dx_{n+i},  \label{4.17}
\end{equation}%
we find
\begin{equation}
i_{X}\Phi _{F^{\ast }}=\Phi _{F^{\ast
}}(X)=X^{3n+i}dx_{i}-X^{i}dx_{3n+i}+X^{2n+i}dx_{n+i}-X^{n+i}dx_{2n+i}.
\label{4.18}
\end{equation}%
Furthermore, the differential of Hamiltonian energy is
\begin{equation}
d\mathbf{H}=\frac{\partial \mathbf{H}}{\partial x_{i}}dx_{i}+\frac{\partial
\mathbf{H}}{\partial x_{n+i}}dx_{n+i}+\frac{\partial \mathbf{H}}{\partial
x_{2n+i}}dx_{2n+i}+\frac{\partial \mathbf{H}}{\partial x_{3n+i}}dx_{3n+i}.
\label{4.19}
\end{equation}%
According to \textbf{Eq.}(\ref{1.1}), \textbf{Eqs. }(\ref{4.18}) and (\ref%
{4.19}) are equaled, we obtain a Hamiltonian vector field given by
\begin{equation}
X=-\frac{\partial \mathbf{H}}{\partial x_{3n+i}}\frac{\partial }{\partial
x_{i}}-\frac{\partial \mathbf{H}}{\partial x_{2n+i}}\frac{\partial }{%
\partial x_{n+i}}+\frac{\partial \mathbf{H}}{\partial x_{n+i}}\frac{\partial
}{\partial x_{2n+i}}+\frac{\partial \mathbf{H}}{\partial x_{i}}\frac{%
\partial }{\partial x_{3n+i}}.  \label{4.20}
\end{equation}

Taking into \textbf{Eq. }(\ref{4.8}), we equal \textbf{Eq. }(\ref{4.10}) and%
\textbf{\ Eq. }(\ref{4.20}), it yields
\begin{equation}
\frac{dx_{i}}{dt}=-\frac{\partial \mathbf{H}}{\partial x_{3n+i}},\text{ }%
\frac{dx_{n+i}}{dt}=-\frac{\partial \mathbf{H}}{\partial x_{2n+i}},\text{ }%
\frac{dx_{2n+i}}{dt}=\frac{\partial \mathbf{H}}{\partial x_{n+i}},\text{ }%
\frac{dx_{3n+i}}{dt}=\frac{\partial \mathbf{H}}{\partial x_{i}}  \label{4.21}
\end{equation}%
Finally, the equations obtained in \textbf{Eq. }(\ref{4.21}) are obtained to
be \textit{Hamiltonian equations} with respect to component $H^{\ast }$ of
almost quaternion structure $V^{\ast }$ on quaternion K\"{a}hler manifold $%
(M,V),$ and then the triple $(M,\Phi _{H^{\ast }},X)$ is a \textit{%
Hamiltonian mechanical system }on quaternion K\"{a}hler manifold $(M,V)$.

\section{Conclusion}

Formalism of Hamiltonian mechanics has intrinsically been described with
taking into account the basis $\{F^{\ast },G^{\ast },H^{\ast }\}$ of almost
quaternion structure $V^{\ast }$ on quaternion K\"{a}hler manifold $(M,V)$.

Hamiltonian models arise to be a very important tool since they present a
simple method to describe the model for mechanical systems. In solving
problems in classical mechanics, the rotational mechanical system will then
be easily usable model.

Since physical phenomena, as well-known, do not take place all over the
space, a new model for dynamic systems on subspaces is needed. Therefore,
equations (\ref{4.11}), (\ref{4.16}) and (\ref{4.21}) are only considered to
be a first step to realize how quaternion geometry has been used in solving
problems in different physical area.

For further research, the Hamiltonian vector fields derived here are
suggested to deal with problems in electrical, magnetical and gravitational
fields of quantum and classical mechanics of physics.

\end{document}